\documentstyle[12pt]{article}

\begin{document}
\begin{titlepage}
\title{\bf Must Quantum Spacetimes Be Euclidean?}
\author{N. Pinto-Neto\thanks{e-mail adress: nelsonpn@lafex.cbpf.br}\\ and\\ E. 
Sergio 
Santini\thanks{e-mail adress: santini@lafex.cbpf.br}\\Centro Brasileiro de
Pesquisas F\'{\i}sicas\\Rua Xavier Sigaud, 150 - Urca\\22290-180,
Rio de Janeiro, RJ\\Brazil}
\maketitle
\newpage
\begin{abstract}
The Bohm-de Broglie interpretation of quantum mechanics is applied to canonical
quantum cosmology. It is shown that, irrespective of any regularization
or choice of factor ordering of the Wheeler-DeWitt equation, the unique
relevant quantum effect which does not break spacetime is the change
of its signature from lorentzian to euclidean. The other
quantum effects are either trivial or break the four-geometry of
spacetime. A Bohm-de Broglie picture of a quantum geometrodynamics is 
constructed,
which allows the investigation of these latter structures.
For instance, it is shown that any real solution of the Wheeler-De Witt
equation yields a generate four-geometry compatible with the strong gravity
limit of General Relativity and the Carroll group.
Due to the more detailed description of quantum geometrodynamics
given by the Bohm-de Broglie interpretation, some new boundary conditions
on solutions of the Wheeler-DeWitt equation must be imposed
in order to preserve consistency of this finer view.

\vspace{0.7cm}
PACS number(s): 98.80.Hw, 04.60.Kz, 04.20.Cv.
\end{abstract}
\end{titlepage}
\section{Introduction}

Almost all physicists believe that quantum mechanics is a universal and
fundamental theory, applicable to any physical system, from which
classical physics can be recovered.  The Universe is, of course, a valid
physical system:  there is a theory, Standard Cosmology, which is able
to describe it in physical terms, and make predictions which can be
confirmed or refuted by observations. In fact, the observations until
now confirm the standard cosmological scenario. Hence, supposing the
universality of quantum mechanics, the Universe itself must be
described by quantum theory, from which we could recover Standard
Cosmology. However, the Copenhaguen interpretation of quantum mechanics
\cite{bohr,hei,von}\footnote{Athough these three authors have different
views from quantum theory, the first emphasizing the indivisibility of
quantum phenomena, the second with his notion of potentiality, and the third
with the concept of quantum states, for all of them the existence of a 
classical domain is crucial. That is why we group their approaches under
the same name ``Copenhaguen interpretation".}, which is the one taught in undergraduate courses
and employed by the majority of physicists in all areas (specially the
von Neumann's approach), cannot be used
in a Quantum Theory of Cosmology. This is because it imposes the existence 
of a classical domain. In von Neumann's view, for instance, the necessity of 
a classical domain comes from the way it solves the
measurement problem (see Ref. \cite{omn} for a good discussion).  In an impulsive measurement of some observable, the wave
function of the observed system plus macroscopic apparatus splits into
many branches which almost do not overlap (in order to be a good
measurement), each one containing the observed system in an eigenstate
of the measured observable, and the pointer of the apparatus pointing
to the respective eigenvalue. However, in the end of the measurement,
we observe only one of these eigenvalues, and the measurement is robust
in the sense that if we repeat it immediately after, we obtain the same
result. So it seems that the wave function collapses, the other
branches disappear. The Copenhaguen interpretation assumes that this
collapse is real.  However, a real collapse cannot be described by the
unitary Schr\"{o}dinger  evolution. Hence, the Copenhaguen interpretation must
assume that there is a fundamental process in a measurement which must
occur outside the quantum world, in a classical domain.
Of course, if we want to quantize the whole Universe,
there is no place for a classical domain outside it, and the Copenhaguen
interpretation cannot be applied. Note that decoherence cannot solve
this problem \cite{muk,zeh}. It can explain why the splitting of the
wave function is given in terms of the pointer basis states, and why we
do not see superpositions of macroscopic objects, due to the effective
diagonalization of the reduced density matrix. However it does not
explain the collapse after the measurement is completed, or why all
but one of the diagonal elements of the density matrix become null
when the measurement is finished. Hence, if
someone insists with the Copenhaguen interpretation, she or he must assume
that quantum theory is not universal, or at least try to improve it by
means of further concepts like in the consistent histories approach
\cite{har}, which is however incomplete until now. Nevertheless, there are
some ways out from this dilemma. We can say that the Schr\"{o}dinger 
evolution is an approximation of a more fundamental non-linear
theory which can accomplish the collapse \cite{rim,pen}, or that the
collapse is effective but not real, in the sense that the other
branches disappear from the observer but do not disappear from
existence. In this second category we can cite the Many-Worlds
Interpretation \cite{eve} and the Bohm-de Broglie Interpretation
\cite{bohm,hol}. In the former, all the possibilities in the splitting
are actually realized. In each branch there is an observer with the
knowledge of the corresponding eigenvalue of this branch, but she or he
is not aware of the other observers and the other possibilities because
the branches do not interfere. In the latter, a point-particle in
configuration space describing the observed system and apparatus is
supposed to exist, independently on any observations.  In the
splitting, this point particle will enter into one of the branches
(which one depends on the initial position of the point particle before
the measurement, which is unknown), and the other branches will be
empty. It can be shown \cite{hol} that the empty waves can neither
interact with other particles, nor with the point particle containing
the apparatus.  Hence, no observer can be aware of the other branches
which are empty.  Again we have an effective but not real collapse (the
empty waves continue to exist), but now with no multiplication of
observers. Of course these interpretations can be used in quantum
cosmology. Schr\"{o}dinger  evolution is always valid, and there is no need
of a classical domain outside the observed system.

In this paper we will foccus on the application of the Bohm-de Broglie
interpretation to quantum cosmology \cite{vink,sht,val,bola1}. In this
approach, the fundamental object of quantum gravity, the geometry of
3-dimensional spacelike hypersurfaces, is supposed to exist independently on
any observation or measurement, as well as its canonical momentum, the extrinsic
curvature of the spacelike hypersurfaces. Its evolution, labeled by some time
parameter, is dictated by a quantum evolution that is different from
the classical one due to the presence of a quantum potential which
appears naturally from the Wheeler-DeWitt equation. This interpretation
has been applied to many minisuperspace models
\cite{vink,bola1,kow,hor,bola2,fab}, obtained by the imposition of
homogeneity of the spacelike hypersurfaces.  The classical limit, the
singularity problem, the cosmological constant problem,
and the time issue have been discussed. For instance, in some of these
papers it was shown that in models involving scalar fields or
radiation, which are nice representatives of the matter content of the
early universe, the singularity can be clearly avoided by quantum
effects. In the Bohm-de Broglie interpretation description, the quantum
potential becomes important near the singularity, yielding a repulsive
quantum force counteracting the gravitational field, avoiding the
singularity and yielding inflation. The classical limit (given by the
limit where the quantum potential becomes negligible with respect to
the classical energy) for large scale factors are usually
attainable, but for some scalar field models it depends on the quantum
state and initial conditions. In fact it is possible to have small
classical universes and large quantum ones \cite{fab}. About the time
issue, it was shown that for any choice of the lapse function the
quantum evolution of the homogeneous hypersurfaces yield the same
four-geometry \cite{bola1}. What remained to be studied is if this fact
remains valid in the full theory, where we are not restricted to
homogeneous spacelike hypersurfaces.  The question is, given an initial
hypersurface with consistent initial conditions, does the evolution of
the initial three-geometry driven by the quantum bohmian dynamics yields
the same four-geometry for any choice of the lapse and shift functions? We
know that this is true if the three-geometry is evolved by the dynamics of
classical General Relativity (GR), but it can be false if the evolving
dynamics is the quantum bohmian one. The purpose of this paper is to
study and answer this question in detail. The idea is to put the
quantum bohmian dynamics in hamiltonian form, and then use strong
results presented in the literature exhibiting the most general
form that a hamiltonian should have in order to form a four-geometry
from the evolution of 3-geometries 
\cite{kuc1}. Our conclusion is that, in general, the quantum bohmian
evolution of the 3-geometries does not yield any four-geometry at all.
Only for very special quantum states a quantum four-geometry can be
obtained, and it must be euclidean. Hence, our answer to the question in
the title is in the affirmative. More important, we arrive at these
conclusions without assuming any regularization and
factor ordering of the Wheeler-DeWitt equation. As we know, the
Wheeler-DeWitt equation involves the application of the product of
local operators on states at the same space point, which is ill defined
\cite{reg}.  Hence we need to regularize it in order to
solve the factor ordering problem, and have a theory free of anomalies
(for some proposals, see Refs \cite{japa1,japa2,kow2}). Our conclusions
are completly independent on these issues. Also, even in the general
case where there is no four-geometry, we can obtain a picture of the
quantum structure yielded by the bohmian dynamics, which is not a
spacetime but something else, as the generate 4-geometries compatible
with the Carroll group \cite{poin}.

This paper is organized as follows: in the next section we 
review the Bohm-de Broglie interpretation of quantum mechanics for
non-relativistic particles and quantum field theory in flat
spacetime. In section 3 we
apply the Bohm-de Broglie interpretation to canonical quantum gravity.
In section 4 we prove that the bohmian evolution of the 3-geometries, irrespective
of any regularization and factor ordering of the Wheeler-DeWitt
equation, can be obtained from a specific hamiltonian, which is of
course different from the classical one.  We then use this
hamiltonian to obtain the main results of our paper concerning the
possibilities of obtaining quantum 4-geometries, and the procedure of
obtaining a picture of other quantum structures.  The classical limits
of these many possibilities are also discussed.  We end with
conclusions and many perspectives for future work.  In the appendix we
present a concrete midisuperspace example of some quantum states which
illustrates the discussion of section 4.

\section{The Bohm-de Broglie Interpretation}

In this section we will review the Bohm-de Broglie interpretation of quantum 
mechanics. We will first show how this interpretation works in
the case of a single particle described by a Schr\"{o}dinger  equation, and
then we will obtain, by analogy, the causal interpretation of quantum field
theory in flat spacetime.

Let us begin with the Bohm-de Broglie interpretation of the
Schr\"{o}dinger  equation describing a single particle. In the coordinate
representation, for a non-relativistic particle with Hamiltonian $
H=p^{2}/2m+V(x),$ the Schr\"{o}dinger  equation is

\begin{equation}
i\hbar \frac{\partial \Psi (x,t)}{\partial t}=\left[ -\frac{\hbar ^{2}}{2m}
\nabla ^{2}+V(x)\right] \Psi (x,t).  \label{bsc}
\end{equation}
We can transform this differential equation over a complex field into a pair
of coupled differential equations over real fields, by writing $
\Psi =A\exp (iS/\hbar )$, where $A$ and $S$ are real functions, and
substituting it into (\ref{bsc}). We obtain the following equations. 
\begin{equation}
\frac{\partial S}{\partial t}+\frac{(\nabla S)^{2}}{2m}+V-\frac{\hbar ^{2}}{2m}\frac{\nabla ^{2}A}{A}=0,  \label{bqp}
\end{equation}
\begin{equation}
\frac{\partial A^{2}}{\partial t}+\nabla \cdot \biggr(A^{2}\frac{\nabla S}{m}\biggl)=0.
\label{bpr}
\end{equation}
The usual probabilistic interpretation, i.e. the Copenhagen interpretation,
understands equation (\ref{bpr}) as a continuity equation for the
probability density $A^{2}$ for finding the particle at position $x$ and
time $t$. All physical information about the system is contained in $A^{2}$,
and the total phase $S$ of the wave function is completely irrelevant. In
this interpretation, nothing is said about $S$ and its evolution equation (\ref{bqp}). 
However, when the term
$\frac{\hbar ^{2}}{2m}\frac{\nabla ^{2}A}{A}$ is negligible, we can interpret
Eqs. (\ref{bqp}) and (\ref{bpr}) as equations for an ensemble of classical particles under the influence
of a classical potential $V$ through the Hamilton-Jacobi equation (\ref{bqp}),
whose probability density distribution in space $A^2(x,t)$ satisfies the
continuity equation (\ref{bpr}), where $\nabla S(x,t) /m$ is the velocity
field $v(x,t)$ of the ensemble of particles. When
$\frac{\hbar ^{2}}{2m}\frac{\nabla ^{2}A}{A}$ is not negligible, we can 
still understand Eq. (\ref{bqp}) as a Hamilton-Jacobi equation 
for an ensemble of
particles. However, their trajectories are no more the classical ones,
due to the presence of the quantum potential term in Eq. (\ref{bqp}).

The Bohm-de Broglie interpretation of
quantum mechanics is based on the {\it two} equations (\ref{bqp}) and
(\ref{bpr}) in the way outlined above, not only on the last one as 
it is the Copenhagen
interpretation. The starting idea is that the position $x$ and momentum
$p$ are always well defined, with the
particle's path being guided by a new field, the quantum field. The field $\Psi $ obeys Schr\"{o}dinger  equation (\ref{bsc}), which can be written as the two real equations
(\ref{bqp}) and (\ref{bpr}). Equation (\ref{bqp}) is 
interpreted as a Hamilton-Jacobi type equation for the quantum particle subjected to
an external potential, which is the classical potential plus the new quantum
potential 
\begin{equation}
Q\equiv -\frac{\hbar ^{2}}{2m}\frac{\nabla ^{2}A}{A}.  \label{qp}
\end{equation}
Once the field $\Psi $, whose effect
on the particle trajectory is through the quantum potential
(\ref{qp}), is obtained from Schr\"{o}dinger equation, we can
also obtain the particle trajectory, $x(t),$ by integrating the differential
equation $p=m\dot{x}=\nabla S(x,t)$, which is called the guidance relation
(a dot means time derivative). Of course, from this differential equation, the
non-classical trajectory $x(t)$ can only be known if the initial
position of the particle is
given. However, we do not know the initial position of the particle
because we do not know how to measure it without disturbances (it is the
hidden variable of the theory). To agree with quantum mechanical
experiments, we have to postulate
that, for a statistical ensemble of particles in the same quantum field 
$\Psi $, 
the probability density distribution of initial positions $x_{0}$
is $P(x_{0},t_0)=A^{2}(x_{0},t=t_0)$.  Equation (\ref{bpr}) guarantees
that $P(x,t)=A^{2}(x,t)$ for all times. In this way, the statistical predictions
of quantum theory in the Bohm-de Broglie interpretation are the same as in the Copenhaguen
interpretation\footnote{It has been shown that under typical chaotic situations,
and only within the Bohm-de Broglie interpretation, a probability distribution
$P \neq A^2$ would rapidly approach the value $P=A^2$ \cite{vig,val2}.
In this case, the probability postulate would be unnecessary, and we could have
situations, in very short time intervals, where this modified Bohm-de Broglie interpretation
would differ from the Copenhaguen interpretation.}.

It is interesting to note that the quantum
potential depends only on the form of $\Psi $, not on its absolute value,
as can be seen from equation (\ref{qp}).
This fact brings home the non-local and contextual character of the quantum
potential\footnote{The non-locality of $Q$ becomes evident when we generalize the causal interpretation to a many
particles system.}. This is a necessary feature because Bell's inequalities
together with Aspect's experiments show that, in general, a quantum theory
must be either non-local or non-ontological. As the Bohm-de Broglie interpretation is
ontological, than it must be non-local, as it is. The non-local and contextual
quantum potential causes the quantum effects.

The function $A$ plays a dual role in the Bohm-de Broglie interpretation: it gives
the quantum potential and the probability density distribution of positions,
but this last role is secondary. If in some model there is no notion of
probability, we can still get information from the system using the
guidance relations. In this case, $A^2$ does not need to be normalizable.
The Bohm-de Broglie interpretation is not, {\em in essence}, a
probabilistic interpretation. It is straightforward to apply it to a single
system.

The classical limit can be obtained in a very simple way. We only have to
find the conditions for having $Q=0$. There is no need to have a classical
domain because this interpretation is ontological. The question on why in a
real measurement we see an effective collapse of the wave function is
answered by noting that, in a measurement, the wave function splits in
a superposition of non-overlapping branches. Hence 
the point particle (representing the particle being measured plus the 
macroscopic apparatus)
will enter into one particular branch, which one depends on the initial conditions, 
and it will be influenced by the quantum
potential related only with this branch, which is the 
only one that is not negligible
in the region where the point particle actually is. The other empty branches continue to exist, but they neither
influence on the point particle nor on any other particle \cite{hol}.
There is an effective but not real collapse. The Schr\"{o}dinger  equation is
always valid.

For quantum fields in flat spacetime, we can apply a similar reasoning.
As an example, take the Schr\"{o}dinger  functional equation for a quantum 
scalar field:    
\begin{equation}
\label{qsf}
i \hbar \frac{\partial \Psi (\phi ,t)}{\partial t} = 
\int d^3x \biggr\{\frac{1}{2}\biggr[-\hbar ^2  
\frac{\delta ^2}{\delta \phi ^2} +
(\nabla \phi)^2\biggl] + U(\phi) \biggl\} \Psi (\phi ,t) .
\end{equation}
Writing again the wave functional as $\Psi = A \exp (iS/\hbar)$, we obtain:
\begin{equation}
\frac{\partial S}{\partial t}+\int d^3x \biggr\{ \frac{1}{2}
\biggr[ \biggr(\frac{\delta S}{\delta \phi}\biggl)^2 +
(\nabla \phi)^2\biggl] + U(\phi) + Q(\phi)\biggl\} = 0 ,
\end{equation}
\begin{equation}
\label{fp}
\frac{\partial A^2}{\partial t}+\int d^3 x \frac{\delta}{\delta \phi}
\biggr(A^2 \frac{\delta S}{\delta \phi}\biggl) = 0 ,
\end{equation}
where $Q(\phi) = -\hbar ^2 \frac{1}{2A} \frac{\delta ^2 A}
{\delta \phi ^2}$ is the corresponding (unregulated) quantum potential.
The first equation is viewed as a modified Hamilton-Jacobi equation
governing the evolution of some initial field configuration through time,
which will be different from the classical one due to the presence
of the quantum potential. The guidance relation is now given by
\begin{equation}
\Pi _{\phi} = \dot{\phi} = \frac{\delta S}{\delta \phi}.
\end{equation}
The second equation is the continuity
equation for the probability density $A^2[\phi(x),t_0]$ of
having the initial field configuration at time $t_0$ given by $\phi(x)$. 

A detailed analysis of the Bohm-de Broglie interpretation of 
quantum field theory is given in Ref. \cite{kal} for the
case of quantum electrodynamics.

\section{The Bohm-de Broglie Interpretation of Canonical Quantum Cosmology}

Let us now apply the Bohm-de Broglie interpretation to canonical quantum 
cosmology. We will
quantize General Relativity Theory (GR) where the matter content is a
minimally coupled scalar field with arbitrary potential. All subsequent
results remain essentially the same for any matter field which couples
uniquely with the metric, not with their derivatives.

The classical hamiltonian of GR with a scalar field is given by:
\begin{equation}
\label{hgr}
H = \int d^3x(N{\cal H}+N^j{\cal H}_j) 
\end{equation}
where
\begin{eqnarray}
\label{h0}
{\cal H} &=& \kappa G_{ijkl}\Pi ^{ij}\Pi ^{kl} + 
\frac{1}{2}h^{-1/2}\Pi ^2 _{\phi}+\nonumber\\
& & + h^{1/2}\biggr[-{\kappa}^{-1}(R^{(3)} - 2\Lambda)+
\frac{1}{2}h^{ij}\partial _i \phi\partial _j \phi+U(\phi)\biggl]\\
\label{hi}
{\cal H}_j &=& -2D_i\Pi ^i_j + \Pi _{\phi} \partial _j \phi .
\end{eqnarray}
In these equations, $h_{ij}$ is the metric of closed 3-dimensional
spacelike hypersurfaces, and $\Pi ^{ij}$ is its canonical momentum
given by
\begin{equation} 
\label{ph}
\Pi ^{ij} = - h^{1/2}(K^{ij}-h^{ij}K) =
G^{ijkl}({\dot{h}}_{kl} -  D _k N_l - D _l N_k ),
\end{equation}
where
\begin{equation} 
K_{ij} = -\frac{1}{2N} ({\dot{h}}_{ij} -  D _i N_j - D _j N_i ) ,
\end{equation} 
is the extrinsic curvature of the hypersurfaces (indices are raisen and lowered
by the 3-metric $h_{ij}$ and its inverse $h^{ij}$). The canonical momentum 
of the scalar field is now
\begin{equation}
\label{pf}
\Pi _{\phi} = \frac{h^{1/2}}{N}\biggr(\dot{\phi}-N^i \partial _i \phi \biggl).
\end{equation}
The quantity $R^{(3)}$ is the
intrinsic curvature of the hypersurfaces and $h$ is the determinant
of $h_{ij}$.
The lapse function $N$ and the shift function $N_j$ are the 
Lagrange multipliers of the super-hamiltonian constraint
${\cal H}\approx 0$ and the super-momentum constraint 
${\cal H}^j \approx 0$,
respectively. They are present due to the invariance of GR under
spacetime coordinate transformations. The quantities $G_{ijkl}$ and
its inverse $G^{ijkl}$ ($G_{ijkl}G^{ijab}=\delta ^{ab}_{kl}$) are
given by
\begin{equation}
\label{300}
G^{ijkl}=\frac{1}{2}h^{1/2}(h^{ik}h^{jl}+h^{il}h^{jk}-2h^{ij}h^{kl}),
\end{equation}
\begin{equation}
\label{301}
G_{ijkl}=\frac{1}{2}h^{-1/2}(h_{ik}h_{jl}+h_{il}h_{jk}-h_{ij}h_{kl}),
\end{equation}
which is called the DeWitt metric. The quantity $D_i$ is the $i$-component
of the covariant derivative operator on the hypersurface, and
$\kappa = 16 \pi G/c^4$.

The classical 4-metric 
\begin{equation}
\label{4g}
ds^{2}=-(N^{2}-N^{i}N_{i})dt^{2}+2N_{i}dx^{i}dt+h_{ij}dx^{i}dx^{j}
\end{equation}
and the scalar field  which are solutions
of the Einstein's equations can be obtained from the Hamilton's equations
of motion
\begin{equation}
\label{hh}
{\dot{h}}_{ij} = \{h_{ij},H\},
\end{equation}
\begin{equation}
\label{hp}
{\dot{\Pi}}^{ij} = \{\Pi ^{ij},H\},
\end{equation}
\begin{equation}
\label{hf}
{\dot{\phi}} = \{\phi,H\},
\end{equation}
\begin{equation}
\label{hpf}
{\dot{\Pi _{\phi}}}= \{\Pi _{\phi},H\},
\end{equation}
for some choice of $N$ and $N^i$, and
if we impose initial conditions compatible with the constraints
\begin{equation}
\label{hh0}
{\cal H} \approx 0 ,
\end{equation}
\begin{equation}
\label{hhi}
{\cal H}_i \approx 0.
\end{equation}
It is a feature of the hamiltonian of GR that the 4-metrics (\ref{4g})
constructed in this way, with the same initial conditions, describe the 
same four-geometry for any choice of $N$ and $N^i$.

The algebra of the constraints close in the following form
(we follow the notation of Ref. \cite{kuc1}):

\begin{eqnarray}\label{algebra}
\{ {\cal H} (x), {\cal H} (x')\}&=&{\cal H}^i(x) {\partial}_i \delta^3(x,x')- 
{\cal H}^i(x'){\partial}_i \delta^3(x',x) \nonumber \\ 
\{{\cal H}_i(x),{\cal H}(x')\}&=&{\cal H}(x) {\partial}_i \delta^3(x,x')  \\ 
\{{\cal H}_i(x),{\cal H}_j(x')\}&=&{\cal H}_i(x) {\partial}_j \delta^3(x,x')+ 
{\cal 
H}_j(x'){\partial}_i \delta^3(x,x') \nonumber  
\end{eqnarray}

To quantize this constrained system, we follow the Dirac quantization 
procedure. The constraints
become conditions imposed on the possible states of the quantum
system, yielding the following quantum equations:
\begin{eqnarray}
\label{smo}
\hat{{\cal H}}_i \mid \Psi  \! > &=& 0 \\
\label{wdw}
\hat{{\cal H}} \mid \Psi  \! > &=& 0 
\end{eqnarray}
In the metric and field representation, the first equation is
\begin{equation}
\label{smo2}
-2 h_{li}D_j\frac{\delta \Psi(h_{ij},\phi)}{\delta h_{lj}} + 
\frac{\delta \Psi(h_{ij},\phi)}{\delta \phi} \partial _i \phi = 0 ,
\end{equation}
which implies that the wave functional $\Psi$ is an invariant under
space coordinate transformations.

The second equation is the Wheeler-DeWitt equation \cite{whe,dew}. Writing
it unregulated in the coordinate representation we get
\begin{equation}
\label{wdw2}
\biggr\{-\hbar ^2\biggr[\kappa G_{ijkl}\frac{\delta}{\delta h_{ij}} 
\frac{\delta}{\delta h_{kl}}
 + \frac{1}{2}h^{-1/2} \frac{\delta ^2}{\delta \phi ^2}\biggl] + 
V\biggl\}\Psi(h_{ij},\phi) = 0 ,
\end{equation}
where $V$ is the classical potential given by
\begin{equation}
\label{v}
V = h^{1/2}\biggr[-{\kappa}^{-1}(R^{(3)} - 2\Lambda)+
\frac{1}{2}h^{ij}\partial _i \phi\partial _j \phi+
U(\phi)\biggl] .
\end{equation}
This equation involves products of local operators at the same
space point, hence it must be regularized. After doing this, one
should find a factor ordering which makes the theory free of anomalies,
in the sense that the commutator of the operator version of the constraints
close in the same way as their respective classical Poisson brackets
(\ref{algebra}). Hence, Eq. (\ref{wdw2}) is only a formal one which must be
worked out \cite{japa1,japa2,kow2}.

Let us now see what is the Bohm-de Broglie interpretation of the solutions of
Eqs. (\ref{smo}) and (\ref{wdw}) in the metric and field representation.
First we write the wave functional in polar form 
$\Psi = A\exp (iS/\hbar )$, where $A$ and $S$ are functionals of
$h_{ij}$ and $\phi$. Substituting it in Eq. (\ref{smo2}), we get two
equations saying that $A$ and $S$ are invariant under general space
coordinate transformations:
\begin{equation}
\label{smos}
-2 h_{li}D_j\frac{\delta S(h_{ij},\phi)}{\delta h_{lj}} + 
\frac{\delta S(h_{ij},\phi)}{\delta \phi} \partial _i \phi = 0 ,
\end{equation}
\begin{equation}
\label{smoa}
-2 h_{li}D_j\frac{\delta A(h_{ij},\phi)}{\delta h_{lj}} + 
\frac{\delta A(h_{ij},\phi)}{\delta \phi} \partial _i \phi = 0 .
\end{equation}

The two equations we obtain for $A$ and $S$ when we substitute
$\Psi = A\exp (iS/\hbar )$ into Eq. (\ref{wdw}) will of course
depend on the factor ordering we choose. However, in any case,
one of the equations will have the form
\begin{equation}
\label{hj}
\kappa G_{ijkl}\frac{\delta S}{\delta h_{ij}} 
\frac{\delta S}{\delta h_{kl}}
 + \frac{1}{2}h^{-1/2} \biggr(\frac{\delta S}{\delta \phi}\biggl)^2
+V+Q=0 ,
\end{equation}
where $V$ is the classical potential given in Eq. (\ref{v}).
Contrary to the other terms in Eq. (\ref{hj}), 
which are already well defined, the precise form of $Q$ depends on the regularization
and factor ordering which are prescribed for the Wheeler-DeWitt equation. 
In the unregulated form given in Eq. (\ref{wdw2}), $Q$ is
\begin{equation}
\label{qp1}
Q = -{\hbar ^2}\frac{1}{A}\biggr(\kappa G_{ijkl}\frac{\delta ^2 A}
{\delta h_{ij} \delta h_{kl}} + \frac{h^{-1/2}}{2} \frac{\delta ^2 A}
{\delta \phi ^2}\biggl) .
\end{equation}
Also, the other equation besides (\ref{hj}) in this case is
\begin{equation}
\label{pr}
\kappa G_{ijkl}\frac{\delta}{\delta h_{ij}}\biggr(A^2
\frac{\delta S}{\delta h_{kl}}\biggl)+\frac{1}{2}h^{-1/2} 
\frac{\delta}{\delta \phi}\biggr(A^2
\frac{\delta S}{\delta \phi}\biggl) = 0 .
\end{equation}

Let us now implement the Bohm-de Broglie interpretation for canonical quantum
gravity.  First of all we note that Eqs. (\ref{smos}) and (\ref{hj}),
which are always valid irrespective of any factor ordering of the
Wheeler-DeWitt equation, are like the Hamilton-Jacobi equations for GR,
suplemented by an extra term $Q$ in the case of Eq. (\ref{hj}), which
we will call the quantum potential. By analogy with the cases of
non-relativistic particle and quantum field theory in flat spacetime, we will
postulate that the 3-metric of spacelike hypersurfaces, the scalar
field, and their canonical momenta always exist, independent on any
observation, and that the evolution of the 3-metric and scalar field
can be obtained from the guidance relations
\begin{equation}
\label{grh}
\Pi ^{ij} = \frac{\delta S(h_{ab},\phi)}{\delta h_{ij}} ,
\end{equation}
\begin{equation}
\label{grf}
\Pi _{\phi} = \frac{\delta S(h_{ij},\phi)}{\delta \phi} ,
\end{equation}
with $\Pi ^{ij}$ and $\Pi _{\phi}$ given by Eqs. (\ref{ph}) and
(\ref{pf}), respectively. Like before, these are first order
differential equations which can be integrated to yield the 3-metric
and scalar field for all values of the $t$ parameter. These solutions
depend on the initial values of the 3-metric and scalar field at some
initial hypersurface.  The evolution of these fields will of course be
different from the classical one due to the presence of the quantum
potential term $Q$ in Eq. (\ref{hj}).  The classical limit is once more
conceptually very simple: it is given by the limit where the quantum
potential $Q$ becomes negligible with respect to the classical energy.
The only difference from the previous cases of the non-relativistic
particle and quantum field theory in flat spacetime is the fact that the
equivalent of Eqs. (\ref{bpr}) and (\ref{fp}) for canonical
quantum gravity, which in the naive ordering is Eq. (\ref{pr}), cannot
be interpreted as a continuity equation for a probabiblity density $A^2$
because of the hyperbolic nature of the DeWitt metric $G_{ijkl}$.
However, even without a notion of probability, which in this case would
mean the probability density distribution for initial values of the
3-metric and scalar field in an initial hypersurface, we can extract a lot of
information from Eq. (\ref{hj}) whatever is the quantum potential $Q$,
as will see now. After we get these results, we will return to this
probability issue in the last section.

First we note that, whatever is the form of the quantum potential $Q$,
it must be a scalar density of weight one. This comes from the
Hamilton-Jacobi equation (\ref{hj}). From this equation we can express
$Q$ as 
\begin{equation}
Q = -\kappa G_{ijkl}\frac{\delta S}{\delta h_{ij}} 
\frac{\delta S}{\delta h_{kl}}
- \frac{1}{2}h^{-1/2} \biggr(\frac{\delta S}{\delta \phi}\biggl)^2 - V .
\end{equation}
As $S$ is an invariant (see Eq. (\ref{smos})), then
$\delta S / \delta h_{ij}$ and $\delta S /\delta \phi$ must be
a second rank tensor density and a scalar density, both of weight one,
respectively. When their products are contracted with $G_{ijkl}$ and
multiplied by $h^{-1/2}$, respectively, they form a scalar density
of weight one. As $V$ is also a scalar density of weight one, then 
$Q$ must also be.
Furthermore, $Q$ must depend only on $h_{ij}$ and $\phi$ because it
comes from the wave functional which depends only on these variables.
Of course it can be non-local (we show an example in the appendix),
i.e., depending on integrals of the fields over the whole space, but
it cannot depend on the momenta.

Now we will investigate the following important problem. From
the guidance relations (\ref{grh}) and  (\ref{grf}) we obtain
the following first order partial differential equations:
\begin{equation} 
\label{hdot}
{\dot{h}}_{ij} =  
2NG_{ijkl}\frac{\delta S}{\delta h_{kl}} + D _i N_j + D _j N_i 
\end{equation}
and
\begin{equation}
\label{fdot}
\dot{\phi}=Nh^{-1/2}\frac{\delta S}{\delta \phi} + N^i \partial _i \phi .
\end{equation}
The question is, given some initial 3-metric and scalar field,
what kind of structure do we obtain when we integrate this equations
in the parameter $t$? Does this structure form a 4-dimensional
geometry with a scalar field for any choice of the lapse and shift
functions? Note that if the functional $S$ were a solution of the
classical Hamilton-Jacobi equation, which does not contain the quantum 
potential term,
then the answer would be in the affirmative because we would be in the
scope of GR. But $S$ is a solution of the {\it modified} Hamilton-Jacobi
equation (\ref{hj}), and we cannot guarantee that this will continue
to be true. We may obtain a complete different structure due to
the quantum effects driven by the quantum potential term
in Eq. (\ref{hj}). To answer this question we will move from
this Hamilton-Jacobi picture of quantum geometrodynamics to a
hamiltonian picture. This is because many strong results concerning
geometrodynamics were obtained in this later picture \cite{kuc1,tei1}.
We will construct a hamiltonian formalism which is consistent with
the guidance relations (\ref{grh}) and (\ref{grf}). It yields the bohmian
trajectories (\ref{hdot}) and (\ref{fdot}) if the guidance relations
are satisfied initially. Once we have this hamiltonian, we can use
well known results in the literature to obtain strong results about
the Bohm-de Broglie view of quantum geometrodynamics.

\section{The Bohm-de Broglie View of Quantum Geometrodynamics}

Examining Eqs. (\ref{smos}) and (\ref{hj}), we can easily guess that
the hamiltonian which generates the bohmian trajectories, once the
guidance relations (\ref{grh}) and (\ref{grf}) are satisfied initially,
should be given by:
\begin{equation}
\label{hq}
H_Q = \int d^3x\biggr[N({\cal H} + Q) + N^i{\cal H}_i\biggl] 
\end{equation}
where we define
\begin{equation}
\label{hq0}
{\cal H}_Q \equiv {\cal H} + Q .
\end{equation}
The quantities ${\cal H}$ and ${\cal H}_i$ are the usual 
GR super-hamiltonian and
super-momentum constraints given by Eqs. (\ref{h0}) and (\ref{hi}).
In fact, the guidance relations (\ref{grh}) and (\ref{grf}) are consistent
with the constraints ${\cal H}_Q \approx 0$ and ${\cal H}_i \approx 0$ 
because $S$ satisfies (\ref{smos}) and (\ref{hj}). Futhermore, they are
conserved by the hamiltonian evolution given by (\ref{hq}). Let us see this in some detail.

First, we write Eqs. (\ref{grh}) and (\ref{grf}) in a sligthly different form 
defining

\begin{equation}
\label{ch}
\Phi^{ij} \equiv \Pi ^{ij} - \frac{\delta S(h_{ab},\phi)}{\delta h_{ij}} 
\approx 0 ,
\end{equation}
and
\begin{equation}
\label{cf}
\Phi _{\phi} \equiv \Pi _{\phi} - \frac{\delta S(h_{ij},\phi)}{\delta \phi}
\approx 0 .
\end{equation}
Let us now calculate $\{\Phi^{ij} (x),{\cal H}_Q (x')\}, \{ \Phi_{\phi} (x),{\cal H}_Q (x')\}, \{\Phi^{ab} (x),{\cal H}_i (x')\}$ and 
$\{ \Phi _{\phi} (x),{\cal H}_i (x')\}$, and see if the guidance relations
(\ref{ch}) and (\ref{cf}), now viewed as constraints, are conserved by the hamiltonian
$H_Q$.

\begin{eqnarray}
\{ {\cal H}_Q (x),\Phi^{ij} (x')\}&=&\kappa \frac{\delta G_{{abcd}}}{\delta h_{ij}'}\Pi^{ab} \Pi^{cd} +  
\frac{1}{2}\frac{\delta h^{-1/2}}{\delta h_{ij}'}\Pi_{\phi}^2 +  
\frac{\delta (V+Q)}{\delta h_{ij}'} \nonumber \\
& & + 2\kappa G_{abcd} \Pi^{ab} 
\frac{\delta^2 S}{\delta h_{cd} \delta h_{ij}'} +  h^{-\frac{1}{2}}
\Pi_{\phi} \frac{\delta^2 S}{\delta \phi \delta h_{ij}'} \nonumber \\ 
&=& \kappa\frac{\delta G_{abcd}}{\delta h_{ij}'}\biggr(\Phi^{ab} \Phi^{cd} +
2\Phi^{ab} \frac{\delta S}{\delta h_{cd}}\biggl)+\frac{1}{2}\frac{\delta h^{-\frac{1}{2}}}{\delta h_{ij}'}\biggr(\Phi _{\phi}^2+  
2 \Phi _{\phi} \frac{\delta S}{\delta \phi}\biggl) \nonumber \\ 
& & + 
2\kappa G_{abcd} \Phi^{ab} \frac{\delta^2 S}{\delta h_{cd} \delta h_{ij}'}+ 
 h^{-\frac{1}{2}} \Phi _{\phi} \frac{\delta S}{\delta \phi \delta h_{ij}'} \nonumber \\
& & +
 \frac{\delta }{\delta h_{ij}'}\biggr[\kappa G_{abcd} \frac{\delta S}{\delta h_{ab}}\frac{\delta S}{\delta h_{cd}} +
 \frac{1}{2}h^{-\frac{1}{2}}\biggr(\frac{\delta S}{\delta \phi}
\biggl)^2 +V+Q\biggl]
\end{eqnarray}
where the primes denote evaluation at $x'$.

The last term is zero because of Eq.(\ref{hj}), and we get

\begin{eqnarray}
\{ {\cal H}_Q (x),\Phi ^{ij} (x')\}&=&\biggr\{\kappa \biggr[-\frac{1}{2}
G_{abcd} h^{ij} \nonumber \\
& & + 
\frac{1}{2} h^{-\frac{1}{2}}(4\delta^{ij}_{ac} h_{bd}- 
\delta^{ij}_{ab}h_{cd}-\delta^{ij}_{cd}h_{ab})\biggl]\biggr(\Phi^{ab} \Phi ^{cd}+2\Phi^{ab} \frac{\delta S}{\delta h_{cd}}\biggl) \nonumber \\
& & -
\frac{1}{4}h^{-\frac{1}{2}} h^{ij} \biggr(\Phi _{\phi}^2 + 
2 \Phi _{\phi} \frac{\delta S}{\delta \phi}\biggl)\biggl\}\delta^3(x,x') \nonumber \\
& &
+ 2 \kappa G_{abcd} \frac{\delta^2 S}{\delta h_{cd} \delta h_{ij}'}\Phi^{ab} + 
h^{-\frac{1}{2}} \frac{\delta^2 S}{\delta \phi \delta h_{ij}'}\Phi _{\phi} \approx 0 .
\end{eqnarray}

In the same way we can prove that

\begin{eqnarray}
{\{\cal H}_Q (x),\Phi _{\phi} (x')\}=2\kappa G_{abcd} \frac{\delta^2 S}{\delta h_{ab} \delta \phi'}\Phi^{cd}+
h^{-\frac{1}{2}}\frac{\delta^2 S}{\delta \phi \delta \phi'}\Phi_\phi\approx 0 ,
\end{eqnarray}
where we have used that the functional derivative of Eq. (\ref{hj}) with respect to $\phi$ is zero. 

For the Poisson brackets involving the supermomentum constraint, as
$S$ is an invariant because it satisfies Eq. (\ref{smos}), then $\Phi^{ij}$
and $\Phi_{\phi}$ are
a second rank tensor density and a scalar density, respectively, both
of weigth one. As ${\cal H}_i$ is the generator of space coordinate
transformations, we get
 
\begin{equation}
\{ {\cal H}_i(x),\Phi ^{ab} (x')\}=-2\delta^{ab}_{ci}\Phi^{cj}(x')
\partial _j \delta ^3(x,x')+
\Phi^{ab}(x) \partial _i \delta ^3(x,x')\approx 0 ,
\end{equation}
and
 
\begin{equation}
{\{\cal H}_i(x),\Phi _{\phi} (x')\}=\Phi _{\phi}
\partial _i \delta ^3(x,x')\approx 0
\end{equation}
 
Combining these results we obtain that

\begin{equation}
\label{cch}
{\dot{\Phi}}^{ij} = \{\Phi ^{ij},H_Q\} \approx  0 ,
\end{equation}
and
\begin{equation}
\label{ccf}
{\dot{\Phi}}_{\phi} = \{\Phi _{\phi},H_Q\} \approx 0 .
\end{equation}
Furthermore, the Poisson brackets of (\ref{ch}) and (\ref{cf})
among themselves are all zero. Finally, the definitions of the momenta
in terms of the velocities 
remain the same as in the classical case because the quantum potential $Q$
does not depend on the momenta:
\begin{equation}
\label{hhq}
{\dot{h}}_{ij} =  \{h_{ij},H_Q\} =  \{h_{ij},H\},
\end{equation}
and
\begin{equation}
\label{hfq}
{\dot{\phi}} = \{\phi,H_Q\} = \{\phi,H\}.
\end{equation}  
Hence we recover (\ref{hdot}) and 
(\ref{fdot}).

We now have a hamiltonian, $H_Q$, which generates the bohmian trajectories
once the guidance relations (\ref{grh}) and (\ref{grf}) are imposed
initially. In the following, we can investigate if the the evolution of the fields
driven by $H_Q$ forms a four-geometry like in classical geometrodynamics.
First we recall a result obtained by Claudio Teitelboim \cite{tei1}.
In this paper, he shows that if the 3-geometries and field configurations
defined on hypersurfaces are evolved by some hamiltonian with the form
\begin{equation}
\label{hg}
\bar{H} = \int d^3x(N\bar{{\cal H}} + N^i\bar{{\cal H}}_i) ,
\end{equation}
and if this evolution can be viewed as the ``motion" of a 3-dimensional
cut in a 4-dimensional spacetime (the 3-geometries can be embedded in
a four-geometry), then the constraints 
$\bar{{\cal H}} \approx 0$ and $\bar{{\cal H}}_i
\approx 0$ must satisfy the following algebra

\begin{eqnarray}
\{ \bar{{\cal H}} (x), \bar{{\cal H}} (x')\}&=&-\epsilon[\bar{{\cal 
H}}^i(x) {\partial}_i \delta^3(x',x)
-  \bar{{\cal H}}^i(x') {\partial}_i \delta^3(x',x)]
\label{algebra1} \\
\{\bar{{\cal H}}_i(x),\bar{{\cal H}}(x')\} &=& \bar{{\cal H}}(x)  
{\partial}_i \delta^3(x,x')  
\label{algebra2} \\
\{\bar{{\cal H}}_i(x),\bar{{\cal H}}_j(x')\} &=& \bar{{\cal H}}_i(x)  
{\partial}_j \delta^3(x,x')- 
\bar{{\cal H}}_j(x') {\partial}_i \delta^3(x,x') 
\label{algebra3}  
\end{eqnarray} 
The constant $\epsilon$ in (\ref{algebra1}) can be $\pm 1$ 
depending if the four-geometry
in which the 3-geometries are embedded is euclidean 
($\epsilon = 1$) or hyperbolic ($\epsilon = -1$).
These are the conditions
for the existence of spacetime.

The above algebra is the same as the algebra (\ref{algebra}) of GR
if we choose $\epsilon = -1$. But the hamiltonian (\ref{hq}) is
different from the hamiltonian of GR only by the presence of
the quantum potential term $Q$ in ${\cal H}_Q$. The Poisson bracket
$\{{\cal H}_i (x),{\cal H}_j (x')\}$ satisfies Eq.
(\ref{algebra3}) because the ${\cal H}_i$ of $H_Q$ defined in Eq.
(\ref{hq}) is the same as in GR. Also 
$\{{\cal H}_i (x),{\cal H}_Q (x')\}$ satisfies Eq. (\ref{algebra2})
because ${\cal H}_i$ is the generator of spatial coordinate tranformations,
and as ${\cal H}_Q$ is a scalar density of weight one (remember that
$Q$ must be a scalar density of weight one), then it must satisfies this
Poisson bracket relation with ${\cal H}_i$. What remains to be verified
is if the Poisson bracket
$\{{\cal H}_Q (x),{\cal H}_Q (x')\}$ closes as in Eq. (\ref{algebra1}).
We now recall the result of Ref. \cite{kuc1}. There it is shown that a 
general super-hamiltonian $\bar{{\cal H}}$ which satisfies Eq.
(\ref{algebra1}), is a scalar density of weight one, whose geometrical
degrees of freedom are given only by the three-metric $h_{ij}$ and its
canonical momentum, and
contains only even powers and no
non-local term in the momenta (together with the other requirements,
these last two conditions are also satisfied
by ${\cal H}_Q$ because it is quadratic in the momenta and the quantum
potential does not contain any non-local term on the momenta), then 
$\bar{{\cal H}}$ must have the following form:

\begin{equation}
\label{h0g}
\bar{{\cal H}} = \kappa G_{ijkl}\Pi ^{ij}\Pi ^{kl} + 
\frac{1}{2}h^{-1/2}\pi ^2 _{\phi} + V_G ,
\end{equation}
where

\begin{equation}
\label{vg}
V_G \equiv -\epsilon h^{1/2}\biggl[-{\kappa}^{-1}(R^{(3)} - 2\bar{\Lambda})+
\frac{1}{2}h^{ij}\partial _i \phi\partial _j \phi+\bar{U}(\phi)\biggr] .
\end{equation}
With this result we can now establish three possible scenarios for the
Bohm-de Broglie quantum geometrodynamics, depending on the form of the quantum 
potential:
\vspace{1.0cm}

{\bf 1) Quantum geometrodynamics evolution forms a four-geometry}

In this case, the Poisson bracket $\{{\cal H}_Q (x),{\cal H}_Q (x')\}$ 
must satisfy Eq. (\ref{algebra1}). Then $Q$ must be such that
$V+Q=V_G$ with $V$ given by (\ref{v}) yielding:
\begin{equation}
\label{q4}
Q = -h^{1/2}\biggr[(\epsilon + 1)\biggr(-{\kappa}^{-1} R^{(3)}+
\frac{1}{2}h^{ij}\partial _i \phi\partial _j \phi\biggl)+
\frac{2}{\kappa}(\epsilon\bar{\Lambda} + \Lambda)+
\epsilon\bar{U}(\phi) + U(\phi)\biggl] .
\end{equation}
Then we have two possibilities:

a) The spacetime is hyperbolic ($\epsilon = -1$)

In this case $Q$ is
\begin{equation}
\label{q4a}
Q = -h^{1/2}\biggr[\frac{2}{\kappa}(-\bar{\Lambda} + \Lambda)
-\bar{U}(\phi) + U(\phi)\biggl] .
\end{equation}
Hence $Q$ is like a classical potential. Its effect is to renormalize the 
cosmological constant and the classical scalar field potential, nothing more.
The quantum geometrodynamics is indistinguishable from the classical one.
It is not necessary to require the classical limit $Q=0$ because $V_G=V+Q$
already may describe the classical universe we live in.

b) The spacetime is euclidean ($\epsilon = 1$)

In this case $Q$ is
\begin{equation}
\label{q4b}
Q = -h^{1/2}\biggr[2\biggr(-{\kappa}^{-1} R^{(3)}+
\frac{1}{2}h^{ij}\partial _i \phi\partial _j \phi\biggl)+
\frac{2}{\kappa}(\bar{\Lambda} + \Lambda)+
\bar{U}(\phi) + U(\phi)\biggl] .
\end{equation}
Now $Q$ not only renormalize the cosmological constant and the 
classical scalar field potential but also change the signature of spacetime.
The total potential $V_G=V+Q$ may describe some era of the early universe
when it had euclidean signature,
but not the present era, when it is hyperbolic. The transition between these
two phases must happen in a hypersurface where $Q=0$, which is the classical
limit. 

We can conclude from these considerations that if a quantum
spacetime exists with different features from the classical observed one, 
then it must be euclidean. In other words, the sole relevant quantum effect which maintains the non-degenerate nature of the four-geometry of spacetime is its
change of signature to a euclidean one. The other quantum effects are either 
irrelevant or break completely the
spacetime structure. This result points in the direction of Ref.
\cite{haw}.
\vspace{1.0cm}

{\bf 2) Quantum geometrodynamics evolution is consistent but does
not form a four-geometry}

In this case, the Poisson bracket $\{{\cal H}_Q (x),{\cal H}_Q (x')\}$ 
does not satisfy Eq. (\ref{algebra1}) but is weakly zero in some other
way. Let us examine some examples.

a) Real solutions of the Wheeler-DeWitt equation.

For real solutions of the Wheeler-DeWitt equation, which is a real
equation, the phase $S$ is null. Then, from Eq. (\ref{hj}), we can
see that $Q=-V$. Hence, the quantum super-hamiltonian  
(\ref{hq0}) will contain only the kinetic term, yielding
\begin{equation}
\label{car}
\{{\cal H}_Q (x),{\cal H}_Q (x')\} = 0.
\end{equation}
This is a strong equality. This case is connected with the strong gravity
limit of GR \cite{tei2,hen,san1}. If we take the limit of big gravitational
constant $G$ (or small speed of light $c$, where we arrive at the Carroll group
\cite{poin}), then the potential in the super-hamiltonian constraint of GR
can be neglected and we arrive at a super-hamiltonian containing only
the kinetic term. The Bohm-de Broglie interpretation is telling us 
that any real solution of the Wheeler-DeWitt equation yields a quantum
geometrodynamics satisfying precisely this strong gravity limit.
The classical limit $Q=0$ in this case implies also that $V=0$.
It should be interesting to investigate further the structure we obtain here.

b) Non-local quantum potentials.

Any non-local quantum potential breaks spacetime but there are some
that may still be consistent. As an example take a quantum potential
of the form
\begin{equation}
\label{non}
Q=\gamma V ,
\end{equation}
where $\gamma$ is a function of the functional $S$ (here comes the
non-locality). In the appendix, we exhibit a wave functional solution of a
midisuperspace model which yields
this type of quantum potential. Let us now calculate $\{{\cal H}_Q (x),{\cal H}_Q (x')\}$:

\begin{eqnarray}
\{ {\cal H}_Q (x), {\cal H}_Q (x')\}&=&\{{\cal H}(x) + Q(x),{\cal H}(x') + Q(x')\} \nonumber \\
 &=&  
\{{\cal H}(x), {\cal H}(x')\}+\{T(x), Q(x')\}+\{Q(x),T(x')\} \nonumber
\end{eqnarray}
where $T$ is the kinetic term of the quantum super-hamiltonian.
Developing the last two terms we get

\begin{eqnarray}
\{ {\cal H}_Q (x), {\cal H}_Q (x')\} &=& \{{\cal H}(x), {\cal H}(x')\} + 
\gamma \{{\cal H}(x), {\cal H}(x')\} \nonumber \\
& & -  
 \frac{d \gamma }{d S}V(x') \biggr[2 \kappa G_{klij}(x)\Pi^{ij}(x)\frac{\delta S}{\delta h_{kl}(x)}+ 
h^{-\frac{1}{2}}\Pi_{\phi}(x)\frac{\delta S}{\delta \phi(x)}\biggl] \nonumber \\
& & + \frac{d \gamma }{d S}V(x)\biggr[2 \kappa G_{klij}(x')\Pi^{ij}(x')\frac{\delta S}{\delta h_{kl}(x')} +  
h^{-\frac{1}{2}}\Pi_{\phi}(x')\frac{\delta S}{\delta \phi(x')}\biggl] 
\nonumber  \\ 
&=& 
(1+\gamma)\{{\cal H}(x), {\cal H}(x')\} \nonumber \\
& & - 
\frac{d \gamma }{d S}V(x')\biggr[2{\cal H}_Q (x) - 2 \kappa G_{klij}(x) \Pi^{ij}(x)\biggr(\Pi^{kl}(x)  
-  \frac{\delta S}{\delta h_{kl}(x)}\biggl) \nonumber \\
& &  -   h^{-\frac{1}{2}}\Pi_{\phi}(x)\biggr(\Pi_{\phi}(x) - \frac{\delta S}{\delta \phi(x)}\biggl)\biggl] \nonumber \\
& &+\frac{d \gamma }{d S}V(x)\biggr[2{\cal H}_Q (x') - 2 \kappa G_{klij}(x')\Pi^{ij}(x')\biggr(\Pi^{kl}(x') - 
\frac{\delta S}{\delta h_{kl}(x')}\biggl) \nonumber \\
& & - h^{-\frac{1}{2}}\Pi_{\phi}(x')\biggr(\Pi_{\phi}(x')-
\frac{\delta S}{\delta \phi(x')}\biggl)\biggl]
\end{eqnarray}
Using the algebra (\ref{algebra}) and the definitions
(\ref{ch}) and (\ref{cf}) we have:

\begin{eqnarray}
\{ {\cal H}_Q (x), {\cal H}_Q (x')\}&=& (1+\gamma)[{\cal H}^i(x) 
{\partial}_i \delta^3(x,x') - {\cal H}^i(x') {\partial}_i \delta^3(x',x)] \nonumber \\
& & - \frac{d \gamma }{d S}V(x')[2{\cal H}_Q (x) - 2 \kappa G_{klij}(x)\Pi^{ij}(x)\Phi^{kl}(x)-  
h^{-\frac{1}{2}}\Pi_{\phi}(x)\Phi_{\phi}(x)] \nonumber \\
& & + \frac{d \gamma }{d S}V(x)[2{\cal H}_Q (x')-  
2 \kappa G_{klij}(x')\Pi^{ij}(x')\Phi^{kl}(x') - h^{-\frac{1}{2}}\Pi_{\phi}(x')\Phi_{\phi}(x')] \nonumber \\
& & \approx 0 
\end{eqnarray}
The rhs in the last expression is weakly zero because it is a combination 
of the constraints and the 
guidance relations
(\ref{ch}) and (\ref{cf}). 
Note that it was very important to use
the guidance relations to close the algebra.
It means that the hamiltonian evolution with the quantum potential
(\ref{non}) is consistent only when restricted to the bohmian trajectories.
For other trajectories, it is inconsitent. Concluding,
when restricted to the bohmian
trajectories, an algebra which does not close in general may close,
as shown in the above example. This is an important remark
on the Bohm-de Broglie interpretation of canonical quantum cosmology, which
sometimes is not noticed\footnote{One could ask if
the guidance relations could not help to close general
algebras as in Eq. (\ref{algebra1}). If this were true, then we could have
more general 
quantum potentials than the ones given in Eq. (\ref{q4}) which would not
break spacetime. However, it can be easily checked that this is not possible.}.

In the examples above, we have explicitly obtained  the 
"structure constants" of the algebra that caracterizes
the  ``pre-four-geometry" generated by $H_Q$ 
i.e., the foam-like structure pointed long time ago in early works 
of J. A. Wheeler \cite{whe,whe2}.

{\bf 3) Quantum geometrodynamical evolution is inconsistent}

In this case, the Poisson bracket $\{{\cal H}_Q (x),{\cal H}_Q (x')\}$ 
is not even weakly zero. The quantum geometrodynamics is inconsistent.
Note that it may be inconsistent even if the theory is free of anomalies
in the sense of section 3. As the Bohm-de Broglie interpretation is a more 
detailed
description of quantum phenomena, its consistency may be more difficult
to achieve. The restriction on the quantum potential along these lines
may yield natural boundary conditions to the Wheeler-DeWitt equation.

\section{Conclusion and Discussions}

The Bohm-de Broglie interpretation of canonical
quantum cosmology yields a quantum geometrodynamical picture where,
in general, and always when the quantum potential is non-local,
spacetime is broken. The 3-geometries evolved under the influence
of a quantum potential do not in general stick together to form a 
non-degenerate 
four-geometry. This is not surprising, as it was antecipated long ago
\cite{whe2}. We obtained this result taking a minimally coupled
scalar field as the matter source of gravitation, but 
it can be generalized to any matter source with non-derivative
couplings with the metric, like Yang-Mills fields. What is nice with 
the Bohm-de Broglie approach is that we can investigate
further what kind of structure is formed, by means of the Poisson
bracket relation (\ref{algebra1}), and the guidance relations
(\ref{hdot}) and (\ref{fdot}). By assuming the existence of 3-geometries,
field configurations, and their momenta, independently on any observations, the
Bohm-de Broglie interpretation allows us to use classical tools, like the 
hamiltonian
formalism, to understand the structure of quantum geometry. If this
information is useful, we do not know. Already in the two-slit 
experiment in non-relativistic quantum mechanics, the Bohm-de Broglie 
interpretation
allows us to say from which slit the particle has passed through:
if it arrive at the upper half of the screen it must have come
from the upper slit, and vice-versa. Such information we do not
have in the many worlds interpretation. However, this information
is useless: we can neither check it nor use it in other experiments.
In canonical quantum cosmology the situation may be the same. 
The Bohm-de Broglie interpretation yields a lot of information about quantum 
geometrodynamics
which we cannot obtain from the many worlds interpretation, 
but this information may 
be useless. However, we cannot answer this question precisely if we
do not investigate further, and the tools are at our disposal.
Furthermore, as the Bohm-de Broglie interpretation is a more detailed 
description
of quantum phenomena, consistency requirements may be more difficult
to achieve. As shown in the last section, we can have inconsistent 
quantum geometrodynamical evolution, even for an anomaly free quantum
theory. This is a nice feature of the Bohm-de Broglie interpretation because
boundary conditions can then be naturally imposed on solutions of the 
Wheeler-DeWitt
equation. Such boundary conditions would be absent in other interpretations.
Also, if we want to be more strict, and impose that quantum geometrodynamics
does not break spacetime, then we will have much more stringent boundary 
conditions. As shown in the previous section, a four-geometry can be attained only 
if the quantum potential
have the the specific form (\ref{q4}).
This is a severe restriction on the solutions of the Wheeler-DeWitt equation
\footnote{These restrictions on the form of the quantum potential do not occur
in minisuperspace models \cite{bola1} because there the hypersurfaces
are restricted to be homogeneous. The only freedom we have is
in the time parametrization of the homogeneous hypersurfaces which
folliate spacetime. There is a single constraint, which of course always
commute with itself irrespective of the quantum potential.}.
In this case, the sole relevant quantum effect will be
a change of signature of spacetime, something pointing towards
Hawking's ideas. Our answer to the question in the title is, then, in the
affirmative.

In the case of consistent quantum geometrodynamical evolution but with no
four-geometry, we have shown that any real solution of the Wheeler-DeWitt equation
yields a structure which is the idealization of the strong gravity limit
of GR. This type of geometry, which is generate, has already been studied 
\cite{san1}. Due to the generality of this picture (it is valid for any
real solution of the Wheeler-DeWitt equation, which is a real equation), it 
deserves
further attention. It may well be that these generate 4-metrics were
the correct quantum gemetrodynamical description of the young universe.
It would be also
interesting to investigate if these structures have a classical limit  
yielding the usual four-geometry of classical cosmology.

For non-local quantum potentials, we have shown that aparently inconsistent
quantum evolutions are in fact consistent if restricted to the bohmian
trajectories satisfying the guidance relations (\ref{grh}) and
(\ref{grf}).
This is a point which is sometimes not taken into account.

Finally, it should be interesting to investigate the connection
between the classical limit and the conditions for inflation and/or
homogeneity and isotropy of the universe. For instance, neglecting
the scalar field, the classical limit of the examples (1-b,2-a,2-b)
in the previous section implies that the initial classical
hypersurface must have a constant scalar curvature, which is
closer to a maximally symmetric initial hypersurface.

We would like to remark that all these results were obtained
without assuming any particular factor ordering and regularization
of the Wheeler-DeWitt equation. Also, we did not use any probabilistic 
interpretation
of the solutions of the Wheeler-DeWitt equation. Hence, it is a quite general
result. However, we would like to make some comments about the
probability issue in quantum cosmology. The Wheeler-DeWitt equation when applied
to a closed universe does not yield a probabilistic interpretation
for their solutions because of its hyperbolic nature. However, it
has been suggested many times \cite{kow,banks,pad,kie,hal} that at
the semiclassical level we can construct a probability measure with
the solutions of the Wheeler-DeWitt equation. Hence, for interpretations where 
probabilities are
essential, the problem of finding a Hilbert space for the solutions
of the Wheeler-DeWitt equation becomes crucial if someone wants to get some
information above the semiclassical level. 
Of course, probabilities are also useful in the Bohm-de Broglie interpretation. 
When we integrate the guidance relations (\ref{hdot}) and (\ref{fdot}), the
initial conditions are arbitrary, and it should be nice to have some
probability distribution on them. However, as we have seen along this paper, 
we can extract a lot of information from the full quantum gravity level
using the Bohm-de Broglie interpretation,
without appealing to any 
probabilistic notion. In this interpretation, probabilities are not
essential. Furthermore, as discussed above, this interpretation
may impose severe boundary conditions from which we could extract the
important results. Hence, we can take the Wheeler-DeWitt equation as it is, without
imposing any probabilistic interpretation at the most fundamental level,
but still obtaining information using the Bohm-de Broglie interpretation,
and then recover probabilities when we reach the semiclassical level.

It would also be important to investigate the Bohm-de Broglie 
interpretation for other
quantum gravitational systems, like black holes. Attempts 
in this direction have been made,
but within spherical symmetry in empty space \cite{japa27}, where we
have only a finite number of degrees of freedom. 
It should be interesting to investigate more general models. These cases are,
however, qualitatively different from quantum closed cosmological models.
There is no problem in thinking of observers outside an ensemble of
black holes. It is quantum mechanics of an open system, with less conceptual
problems of interpretation.

The conclusions of this paper are of course limited by many strong
assumptions we have tacitly made, as supposing that a continuous three-geometry 
exists at the quantum level (quantum effects could also destroy it), or
the validity of quantization of standard GR, forgetting other developments
like string theory. However, even if this approach is not the apropriate one,
it is nice to see how far we can go with the Bohm-de Broglie interpretation,
even in such incomplete stage of canonical quantum gravity.
It seems that the Bohm-de Broglie interpretation may at least be 
regarded as a nice ``gauge" \cite{bra} to be used in quantum cosmology,
as, probably, it will prove harder, or even impossible,
to reach the detailed conclusions of this paper using other interpretations.
However, if the finer view of the Bohm-de Broglie interpretation of quantum
cosmology can yield useful information in the form of observational effects, 
then we will have means to
decide between interpretations, something that will be very important not
only for quantum cosmology, but for quantum theory itself.
\vspace{1.0cm}

{\bf ACKNOWLEDGEMENTS}

We would like to thank Brandon Carter and the Cosmology Group of CBPF for  useful discussions. Our special warm thanks are to  Karel Kucha$\check{\mbox{r}}$ for many helpful and lucid comments on our work
while he was visiting our institute.
We would also like to thank CNPq of Brazil for financial support.
\vspace{1.0cm}

{\bf Appendix: A Non-local Quantum Potential in the Bohm-de Broglie Interpretation of a Quantum Spherical Spacetime}

We present here an example where a quantum potential of the type 
(2-b) is obtained. 
It is an spherically symmetric midisuperspace model with an electromagnetic 
field. A factor-ordering
is proposed just to yield an exact solution of the Wheeler-DeWitt equation, 
which is not regularized\footnote{In fact, the model studied here can be reduced
to a minisuperspace model, with a finite number of degrees of freedom
\cite{kuchar}. However, our goal with this appendix is just to argue that
quantum potentials of the type (2-b) are not so difficult to appear
in canonical quantum gravity.
They may be obtained already from spherically symmetric Wheeler-DeWitt 
equations, as we will see.}.

We start from the ADM decomposition of the general spherically symmetric electrovacuum 
spacetime metric on the manifold $R\times R \times S^2$:

\begin{equation} ds^2=-N^2dt^2+\Lambda^2(dr+N^{r}dt)^2+R^2d\Omega^2 \end{equation}
where $N$, $N^{r}$ are the lapse and shift functions respectively (both dependent on $r$ and $t$),
and $d\Omega$ denotes the standard line element on $S^2$.
The electromagnetic potential is taken to be described by the spherically symmetric one-form:

\begin{equation} dA=\Gamma(r,t)dr+\Phi(r,t)dt \end{equation}
The ADM action in midisuperspace, 
after integrating over the two sphere, reads ($c\equiv 1$)

\begin{eqnarray}S=\int dt \int dr \frac{1}{2N} \biggr\{ \frac{1}{G}\biggr[N^2\Lambda-\Lambda \dot{R}^2
+2\frac{N^2 R \Lambda' R'}{\Lambda^2}-2\frac{N^2 R R''}{\Lambda}- \nonumber \\ \Lambda (N^{r})^2 R'^2 
+2N^{r}\dot{R}(\Lambda R)'-2 R N^{r} R'(\Lambda N^{r})'+ 2 R \Lambda (N^{r})' \dot{R} \nonumber \\ 
+ 2 R N^{r} \dot{\Lambda} R'-2 R \dot{\Lambda} \dot{R}- 
N^2 \frac{R'^2}{\Lambda}\biggl] + 
\frac{R^2}{\Lambda}(\dot{\Gamma}-\Phi')^2 \biggl\} ,
\end{eqnarray}
where a prime denotes a derivative in $r$.
Varying the action with respect to $N$ and $N^{r}$ leads to the 
super-hamiltonian 
and the super-momentum constraints \cite{kuchar,louko}

\begin{equation} {\it {\cal H}}\equiv \frac{G}{2}\frac{\Lambda P_{\Lambda}^2}{R^2}-
G\frac{P_{\Lambda}P_{R}}{R}
+\frac{V_g}{G}+ \frac{\Lambda P_{\Gamma}^2}{2R^2}\approx 0 , 
\end{equation}
and

\begin{equation}\label{mc} {\it {\cal H}_{r}}\equiv P_{R}R'-\Lambda P_{\Lambda}'\approx 0 , 
\end{equation}
where

\begin{equation}\frac{V_g}{G}=\frac{RR''}{\Lambda}-
\frac{RR'\Lambda'}{\Lambda^2}+\frac{R'^2}{2\Lambda}-\frac{\Lambda}{2}
\end{equation}
Varying the action with respect to the Lagrange multiplier $\phi$ leads to

\begin{equation}
\label{E} 
P_{\Gamma}' \approx 0 .
\end{equation}
Boundary conditions for all fields are assumed to hold such that all integrals are well defined,
and such that the classical spacetime metric is nondegenerate \cite{louko}.

We perform the quantization in the standard formal manner. All constraints act on wave functionals $\Psi[\Lambda(r),R(r),\Gamma(r)]$. The electromagnetic constraint (\ref{E}) is solved \cite{brotz} 
by $ \Psi= f(\int_{\infty}^{\infty}\Gamma dr)\psi[\Lambda(r),R(r)]$,
where $f$ is an arbitrary differentiable function.

From the hamiltonian constraint, we obtain the Wheeler-DeWitt equation with a particular factor ordering:

\begin{equation}
\biggr(-\frac{G\hbar^2 \Lambda}{2R^2}F\frac{\delta}{\delta\Lambda}F^{-1}
\frac{\delta}{\delta\Lambda}
+\frac{G\hbar^2}{R}F \frac{\delta}{\delta R} F^{-1} \frac{\delta}{\delta \Lambda}+\frac{V_g}{G}-
 \frac{\hbar^2 \Lambda \delta^2}{2R^2\delta \Gamma^2}\biggl)\Psi=0 
\end{equation}
where

\begin{equation} F\equiv R \sqrt{\biggr(\frac{R'}{\Lambda}{\biggl)}^2+\frac{2m}{R}-\frac{q^2}{R^2}-1} 
\end{equation}
Our solution has the form \cite{brotz}

\begin{equation} 
\Psi=e^{\frac{iq}{\hbar}\int_{-\infty}^{\infty}\Gamma dr}
\psi[\Lambda(r),R(r)] , 
\end{equation}
where the gravitational and electromagnetic degrees of freedom are 
separated, and $\psi$ satisfies a reduced Wheeler-DeWitt equation

\begin{equation}
\label{wd} \biggr(-\frac{G\hbar^2 \Lambda}{2R^2}F\frac{\delta}{\delta\Lambda}F^{-1}
\frac{\delta}{\delta\Lambda}
+\frac{G\hbar^2}{R}F \frac{\delta}{\delta R} F^{-1} \frac{\delta}{\delta \Lambda}+\frac{V_g}{G}+
\frac{\Lambda q^2}{2 R^2}\biggl)\psi=0 .
\end{equation}
We are using this particular factor ordering because in this case there is  an exact solution 
known in the literature \cite{brotz}, which reads

\begin{equation}
\label{psi1} \psi=\exp{\frac{iS_{0}}{\hbar}} , 
\end{equation}
where

\begin{equation}
\label{so} 
S_{0}= G^{-1}\int_{-\infty}^{\infty}dr \biggr\{\Lambda F -
 \frac{1}{2} RR'\ln{\frac{\frac{R'}{\Lambda}+\frac{F}{R}}{\frac{R'}{\Lambda}
 -\frac{F}{R}}}\biggl\} 
\end{equation}
It is easy to see that, as the Wheeler-DeWitt equation is real, the complex conjugate of the wave 
functional (\ref{psi1}) is another exact solution of it.
Hence, we have two independent exact solutions of the Wheeler-DeWitt equation, and because of its linearity, 
any linear superposition of them will also be a solution :

\begin{equation}
\label{psi} \psi=a\exp{\biggr(\frac{iS_{0}}{\hbar}\biggl)}+
b\exp{\biggr(\frac{-iS_{0}}{\hbar}\biggl)} .
\end{equation}
Writing the wave functional in polar form

\begin{equation}
\psi=A\exp{\frac{iS}{\hbar}} , 
\end{equation}
and replacing it in the Wheeler-DeWitt equation (\ref{wd}), we obtain two equations. 
One of them reads

\begin{equation}
\label{cwdw}
\frac{G\Lambda}{2R^2}(\frac{\delta S}{\delta\Lambda})^2-\frac{G}{R}\frac{\delta S}{\delta\Lambda}
 \frac{\delta S}{\delta R} + V + Q=0 ,
\end{equation}
where $V$ stands for  the classical potential

\begin{equation}
V \equiv \frac{V_g}{G}+
\frac{\Lambda q^2}{2 R^2} , 
\end{equation}
and $Q$ is the quantum potential

\begin{equation}\label{Pq}
Q=\frac{G\hbar^2}{A R}\biggr(-\frac{\Lambda \delta^2 A}{2R\delta \Lambda^2} +\frac{\delta^2 A}{\delta R \delta \Lambda}+
\biggr(-\frac{1}{F}\frac{\delta F}{\delta R} +\frac{\Lambda}{2 R F}\frac{\delta F}{\delta \Lambda}\biggl)\frac{\delta A}{\delta\Lambda}\biggl) .
\end{equation}
For the wave functional (\ref{psi}), the quantum potential (\ref{Pq}) reads

\begin{equation}
Q= \gamma  V ,
\end{equation}
where $V$ is the classical potential, and the factor $\gamma$ is given by

\begin{equation}
\label{pq} \gamma=-4\biggr\{ \biggr(\frac{ab}{A^2}\biggl)^2 \sin^2{\biggr(\frac{2 S_{0}}{\hbar}\biggl)} +
\frac{ab}{A^2}\cos{\biggr(\frac{2 S_{0}}{\hbar}\biggl)} \biggl\} .
\end{equation}
As the phase functional $S$ is given in terms of $S_{0}$ by
$S=\frac{\hbar}{2i}\ln(\frac{\psi}{\psi^*})$, where $\psi$ is a function
of $S_0$ only, then 

\begin{equation}
\gamma=\gamma(S) .
\end{equation}
Hence we have found a concrete example of a quantum potential of the
type studied in (2-b),
where the constraint's algebra close with  new ``structure constants", 
breaking the four-geometry of spacetime.

\end{document}